\documentclass[10pt, conference]{IEEEtran}

\newcommand{\takeaway}[1]{\noindent{\textit{\textbf{Takeaway:}}} \textit{#1}}

\newcommand{\ie}{i.e., }

\usepackage{times}
\usepackage[utf8]{inputenc}
\usepackage[T1]{fontenc}
\usepackage[nolist]{acronym}
\usepackage[colorlinks=true,allcolors=blue]{hyperref}
\usepackage{url}
\usepackage{pifont}
\usepackage{multirow}
\usepackage{siunitx}
\usepackage{array,booktabs}

\usepackage[caption=false]{subfig}
\usepackage{tabularx}
\usepackage{cite}
\usepackage{amsmath,amssymb,amsfonts}
\usepackage{algorithmic}
\usepackage{graphicx}
\usepackage{textcomp}
\usepackage[usenames,dvipsnames,svgnames,table]{xcolor}
\usepackage{color,soul}
\usepackage{balance}
\usepackage{longtable}
\usepackage[caption=false]{subfig}

\usepackage{tikz}
\newcommand\copyrighttext{%
  \footnotesize \textcopyright~IFIP, 2022. This is the author's version of the work. It is posted here by permission of IFIP for your personal use. Not for redistribution. The definitive version was published in 2022 IFIP Networking Conference, \url{https://doi.org/10.23919/IFIPNetworking55013.2022.9829773}}
	\newcommand\copyrightnotice{%
	\begin{tikzpicture}[remember picture,overlay]
	\node[anchor=south,yshift=10pt] at (current page.south) {\fbox{\parbox{\dimexpr\textwidth-\fboxsep-\fboxrule\relax}{\copyrighttext}}};
	\end{tikzpicture}%
}

\begin{acronym}
    \acro{pep}[PEP]{Performance Enhancing Proxy}
    \acro{loops}[LOOPS]{Local Optimizations on Path Segments}
    \acro{rto}[RTO]{Retransmission timeout}
    \acro{fec}[FEC]{Forward Error Correction}
    \acro{holb}[HOL blocking]{Head-of-line blocking}
    \acro{quic}[QUIC]{Quick UDP Internet Connections}
    \acro{vpn}[VPN]{Virtual Private Network}
    \acro{masque}[MASQUE]{Multiplexed Application Substrate over QUIC Encryption}
    \acro{ietf}[IETF]{Internet Engineering Task Force}
    \acro{wg}[WG]{Working Group}
    \acro{tcpm}[TCPM]{TCP Maintenance and Minor Extensions}
    \acro{tcp}[TCP]{Transmission Control Protocol}
    \acro{bbr}[BBR]{Bottleneck Bandwidth and Round-trip propagation time}
    \acro{satcom}[SATCOM]{Satellite Communication}
    \acro{cca}[CCA]{Congestion Control Algorithm}
    \acro{iw}[IW]{Initial Window}
    \acro{alpn}[ALPN]{Application-Layer Protocol Negotiation}
    \acro{os}[OS]{Operating System}
    \acro{bdp}[BDP]{Bandwidth-Delay Product}
\end{acronym}

\acrodefplural{vpn}[VPNs]{Virtual Private Networks}
\acrodefplural{pep}[PEPs]{Performance Enhancing Proxies}
\acrodefplural{cca}[CCAs]{Congestion Control Algorithms}

\begin{document}

\title{Exploring Proxying QUIC and HTTP/3 \\for Satellite Communication\vspace{-0em}}

\author{Mike Kosek$^\star$, Hendrik Cech$^\star$, Vaibhav Bajpai$^\dagger$, Jörg Ott$^\star$\\
$^\star$Technical University of Munich, Germany\\
\texttt{[kosek | cech | ott]@in.tum.de}\\
$^\dagger$CISPA Helmholtz Center for Information Security\\
\texttt{bajpai@cispa.de}
\vspace{-0.5em}
}

\maketitle


\begin{abstract}
  Low-Earth Orbit satellites have gained momentum to provide Internet connectivity, augmenting those
  in the long-established geostationary orbits. At the same time, QUIC has been developed as the
  new transport protocol for the web.
  While QUIC traffic is fully encrypted, intermediaries such as performance enhancing proxies (PEPs) -- in the past essential for Internet over satellite performance -- can no longer tamper with and optimize transport connections.
  In this paper, we present a satellite emulation testbed and use it to compare QUIC and TCP as well as HTTP/3 and HTTP/1.1 with and without minimal PEP functionality.
  Evaluating \textit{goodput} over time, we find that the slow start threshold is reached up to 2s faster for QUIC PEP in comparison to QUIC Non-PEP.
  Moreover, we find that HTTP/3 and HTTP/3-PEP outperform HTTP/1.1 and HTTP/1.1-PEP in multiple \textit{web performance} scenarios, where HTTP/3-PEP improves over HTTP/3 for \emph{Page Load Time} by over 7s in edge cases.
  Hence, our findings hint that these performance gains may warrant exploring PEPs for QUIC.
  \copyrightnotice

\end{abstract}

{
  \let\thefootnote\relax\footnotetext{ISBN 978-3-903176-48-5\textcopyright~2022 IFIP}
}



\begin{figure*}[t]
	\centering
	\includegraphics[width=0.9\linewidth,trim=13.4cm 10.8cm 13.4cm 10.8cm, clip]{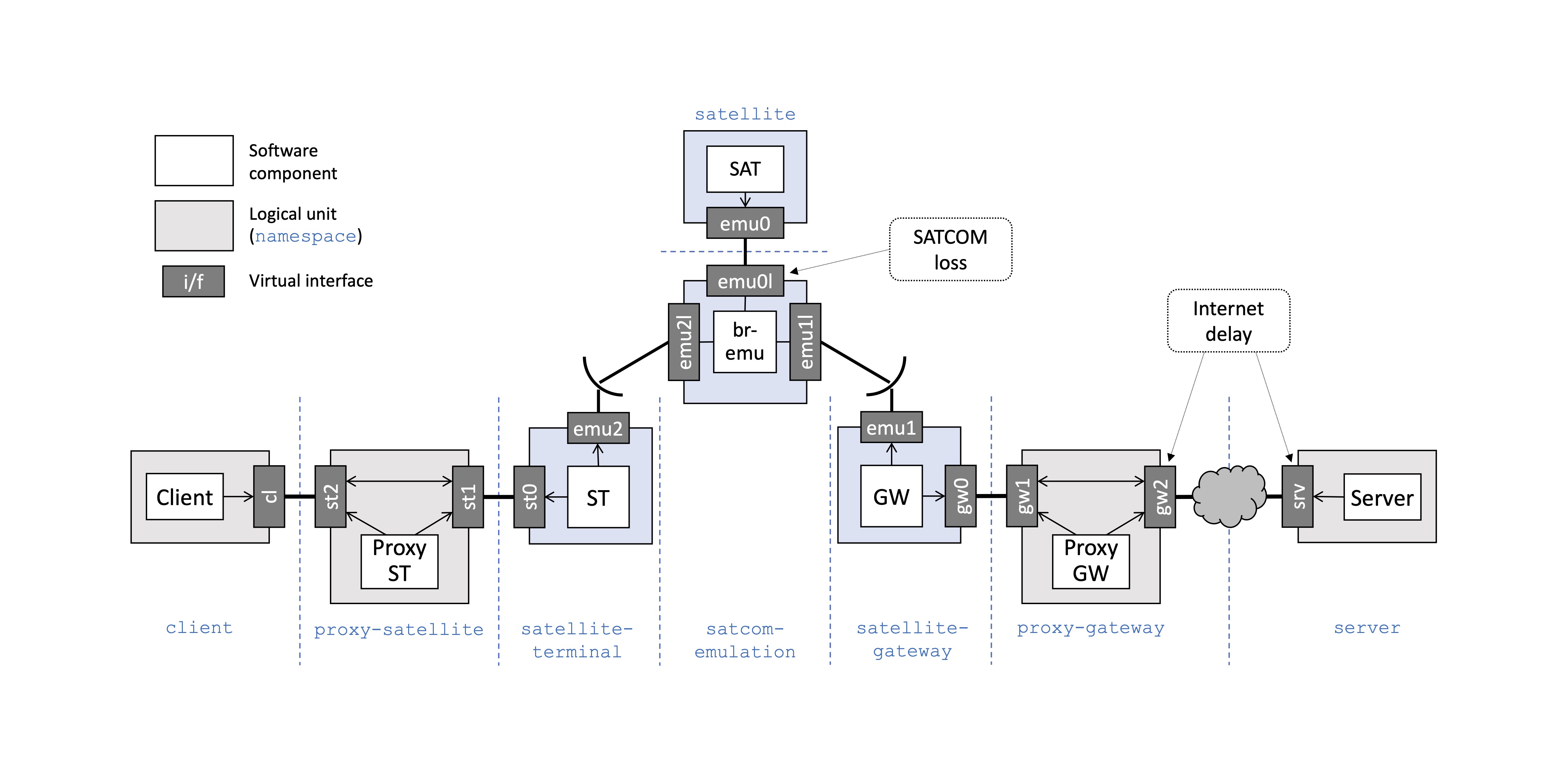}
	\caption{
	The SATCOM emulation testbed comprises eight logical units based on Linux network namespaces depicted as light shaded rectangles, where the blue shaded ones make up the OpenSAND SATCOM components (\emph{ST, SAT, GW}).
	\emph{Client} and \emph{Server} are the measurement endpoints and two PEPs (\emph{Proxy-ST, Proxy-GW}) may be included in the path or bypassed.
	To induce \emph{SATCOM loss}, the forwarding properties of the indicated interface are modified by \texttt{netem} emulation components.
	In addition, \texttt{netem} is also used to create \emph{Internet delay} between the satellite ground station hosting the \emph{Proxy-GW} and the target server.
	Arrows from components to interfaces indicate network access through these interfaces, while double-sided arrows represent routing between networks.
	Bold lines connect two interfaces and thin lines indicate the affiliation with a bridge.
	} 
	\vspace{-1em}
	\label{fig:satcom-measurement-testbed}
\end{figure*}

\section{Introduction}
\label{sec:introduction}
Internet access via satellite has experienced a revival with the
advent of Starlink by SpaceX \cite{starlink}, Kuiper by Amazon
\cite{kuiper}, Oneweb \cite{oneweb}, and Telesat \cite{telesat}, which
leverage massive satellite constellations in low earth orbits (LEO) at
 300--2000\,km altitude.
This is contrast to early services, such as Hughes DirecTV and Viasat
Connexion for aircraft, among others, that mostly used geostationary
(GEO) satellites at an altitude of 35,785km to provide Internet
connectivity in the late 1990s and (early) 2000s, with Iridium
\cite{iridium} as one notable exception that has also relied on LEO
satellites.

Using GEO satellites, such early Internet access services have experienced a
one-way delay of around 250ms for the satellite hop.  This called
for introducing intermediaries, often dubbed \textit{performance
  enhancing proxies (PEPs)}~\cite{rfc3135}, in order to improve (interactive)
content access. PEPs may generally provide two classes of
functionality:
%
\textbf{I.} Transport layer connection splitting and other transport
layer optimizations aim at decoupling the congestion and error control
loops of different path segments, usually isolating the ``challenged''
(\ie typically satellite) link \cite{rfc3135}.  These functions can be
performed even if higher layer security protocols (e.g., TLS) are used
as TCP headers are in the clear and allow the manipulation of connections.
\textbf{II.} Application layer functions such as HTTP prefetching and
content caching aim at reducing the impact of high RTTs on application layer protocols. These functions need access to
the application data and thus won't work unless secure connections are
terminated at an intermediary. While, for the regular Internet,
Content Delivery Networks (CDNs) may help by replicating services and
contents close to the user, this is generally not applicable when the
challenged link or path segment includes the last hop.

Today, the advent of QUIC \cite{rfc9000,rfc9001,rfc9002} as a secure
transport protocol that also encrypts control information prevents
intermediaries from accessing header fields, and thus challenges the
implementation and use of the transport layer PEP functionality (I.).
Yet, connection splitting remains possible if proxies do not operate
transparently but are explicitly included in QUIC connection setup
as, e.g., discussed in the MASQUE WG of the IETF~\cite{masque}.
But would QUIC PEPs actually make a difference -- for both traditional
GEO-based satellite services and for the recent LEO-based ones?

This question arises for (at least) two reasons:
(1) QUIC integrates transport setup and TLS 1.3 security and thus
reduces the number of round-trips required for connection
establishment, possibly even eliminating them entirely with 0-RTT
setup for recurring connections to the same server.
(2) QUIC supports stream multiplexing within a connection without
head-of-line blocking so that only a single connection per server is 
needed, eliminating repeated setup costs.

In this paper, we seek to explore if these two performance improvements
built into QUIC by design are sufficient to offset the need for PEPs
in satellite networks.  Since end-to-end encryption using TLS/TCP or
QUIC rules out application layer PEP operations (II.), we focus on I., \ie
transport layer optimizations.
To this end, we make two main contributions:

(1) We present a satellite communication emulation testbed 
(§~\ref{sec:satcom-measurement-testbed}) which enables reproducible 
measurements over satellite networks by using our specifically designed QUIC PEP as well 
as QUIC performance measurement implementations.
%

(2) We carry out an extensive emulation study
(§~\ref{sec:quic-proxies}) assuming propagation delays of LEO and GEO
satellites and explore various combinations of link characteristics.
We report on connection \textit{goodput} over time and, as initial indicators on web performance, \textit{Response Start}, \textit{First Contentful Paint}, and \textit{Page Load Time}.

We compare QUIC vs. TCP, and HTTP/3 (which uses QUIC) vs. HTTP/1.1 (which uses TCP), with and without PEPs, representing the past and future web.
Evaluating \textit{goodput} over time, we find that the slow start threshold is reached up to 2s faster for QUIC PEP in comparison to QUIC Non-PEP, where the improvements are more pronounced on connections with higher RTTs.
Moreover, we find that HTTP/3 and HTTP/3-PEP outperform HTTP/1.1 and HTTP/1.1-PEP in multiple scenarios which we attribute to QUIC's multiplexing capabilities. 
In addition, HTTP/3-PEP improves over HTTP/3 for \emph{Page Load Time} in GEO orbits: With a reduction of $\sim$330ms for real world conditions, and over 7s in edge cases, we observe a benefit of PEPs for QUIC connections.

In order to enable the reproduction of our findings, we make the developed tools, the raw data of our measurements, as well as the analysis scripts and supplementary files publicly available\footnote{https://github.com/kosekmi/2022-ifip-nw-quic-proxies}.
We note that our current QUIC PEP evaluation assumes a simplistic setup in which the PEP terminates a QUIC connection and then relays the user data, exposing it to the proxy.
While other designs are conceivable, this choice does not affect the purpose of our measurements: understanding if QUIC could benefit from PEPs.
While we discuss this and further limitations of our current work in §~\ref{sec:limitations}, related work will be detailed in §~\ref{sec:related-work}, and §~\ref{sec:conclusion} concludes the paper.

\section{SATCOM Emulation Testbed}
\label{sec:satcom-measurement-testbed}

The \emph{\ac{satcom} emulation testbed} enables reproducible transport as well as application layer measurements over \ac{satcom} networks, leveraging OpenSAND~\cite{opensand} for the emulation of the satellite components.
OpenSAND is an established open-source tool for the emulation of SATCOM networks, featuring link-layer emulation based on the DVB-RCS2 and DVB-S2 standards~\cite{dvbs2}.
While OpenSAND shows a high degree of accuracy~\cite{trustable_sat_emulations} and is widely used~\cite{qpep, opensand_ref_1, opensand_ref_2, trustable_sat_emulations}, its complex setup and parametrization is considered a major obstacle~\cite{trustable_sat_emulations}.
We therefore abstract the parametrization of the \emph{\ac{satcom} emulation testbed}, creating a controlled emulation environment executed on a single Linux system using different \emph{scenarios}.
A \emph{scenario} represents the combination of different testbed (e.g., delay, loss, attenuation) as well as transport layer (e.g., congestion control, initial window) parameters used by the emulation.
Each \emph{scenario} can be run a specified number of times, where the testbed is gracefully shut down and newly started on every emulation run to rule out any influence of previous runs to a subsequent one.
Within each \emph{scenario}, multiple transport and application layer \emph{measurement types} are performed: using QUIC, TCP, HTTP/3, as well as HTTP/1.1, each with and without the aid of a \ac{pep}.
By combining multiple \emph{scenarios} within one emulation configuration, the automated visualization enables a holistic view and detailed analysis of transport and application layer performance over all measured \emph{scenarios}.

In this section, §~\ref{sub:testbed-design} details the \emph{SATCOM emulation testbed design}, followed by a comprehensive description of the \emph{scenario} parameters in §~\ref{sub:scenarios}.
While §~\ref{sub:measurement-types} introduces the \emph{measurement types}, §~\ref{sub:quic-pep} presents the \emph{QUIC PEP} implementation in order to proxy QUIC connections.
Following the introduction of the \emph{QUIC performance measurement tool} in §~\ref{sub:qperf}, we conclude with a \emph{validation} of the emulation testbed in §~\ref{sub:validation}.

\subsection{Design}
\label{sub:testbed-design}

Fig.~\ref{fig:satcom-measurement-testbed} depicts the SATCOM emulation testbed design.
The emulation comprises eight logical units based on Linux network namespaces depicted als shaded rectangles.
On the left-hand side, the \emph{Client} is deployed, running the client-side measurement tools for the respective \emph{measurement type}.
The \texttt{client} namespace is connected to the \texttt{proxy-satellite} namespace, where, depending on the \emph{measurement type}, the \emph{Proxy-ST} proxies the data, or an IP route will forward the data to the \texttt{satellite-terminal} namespace.
For the emulation of the satellite components (blue shaded rectangles), we leverage OpenSAND.
Within the \texttt{satellite-terminal} namespace, the interface \texttt{st0} forwards the packets to the emulated \emph{satellite terminal ST}.
\emph{ST} encapsulates the packets into DVB-RCS2 RLE (Return Link Encaspulation) frames and forwards the data to the \texttt{satcom-emulation} namespace which interconnects the three OpenSAND components \emph{ST}, \emph{SAT}, and \emph{GW} (blue shaded rectangles) via the \texttt{bridge}-interface \texttt{br-emu}.
The \emph{SAT} component performs the satellite emulation (e.g., packet delay, signal attenuation), where loss is also emulated on the satellite connection (interface \texttt{emu0l}) using \texttt{netem}.
Following the satellite emulation, packets are sent to the \texttt{satellite-gateway} namespace, where the \emph{satellite gateway GW} de-encapsulates the DVB-RCS2 RLE frames, and forwards the data to the \texttt{proxy-gateway} namespace.
Depending on the \emph{measurement type}, the \emph{Proxy-GW} proxies the packets, or an IP route forwards the data.
Using either option, the \texttt{gw2} interface subsequently emulates the one-way delay between the \texttt{proxy-gateway} and the \emph{Server} using \texttt{netem}, i.e., the delay between the satellite ground station hosting the \emph{Proxy-GW} and the target server.
As a final step, the \emph{Server} runs the server-side measurement tools for the respective \emph{measurement type} and replies to the measurement requests initiated by the \emph{Client}.
While the communication flow in the opposite direction, i.e., from \emph{Server} to \emph{Client}, is mainly identical, the \emph{Internet delay} is added on \texttt{srv} instead of \texttt{gw2}, the encapsulation/de-encapsulation of \emph{Proxy-GW} and \emph{Proxy-ST} are interchanged, and the packets are encapsulated using DVB-S2 GSE (Generic Stream Encapsulation) instead of DVB-RCS2 RLE.

The depicted \emph{SATCOM testbed design} represents a typical \ac{satcom} use-case: A client connects via a local network to a \emph{satellite-terminal}, which communicates via a \emph{satellite} to a \emph{satellite-gateway}, from where a connection to a server is established.
In this use-case, the \emph{satellite-terminal} represents the access router, where the \emph{satellite-gateway} represents the ground station -- both operated by the SATCOM network provider.
Moreover, both \emph{Proxy-ST} and \emph{Proxy-GW} are also operated by the SATCOM network provider in order to optimize the SATCOM connections using \acp{pep}.

\subsection{Scenarios}
\label{sub:scenarios}
A \emph{scenario} represents the combination of different testbed and transport layer parameters used by the emulation, which are presented in Tab.~\ref{tab:scenario-parameters} and detailed in the following.

\textbf{Testbed parameters.}
The \emph{Internet delay} states the one-way delay between the satellite ground station and the target server.
In contrast, the \emph{SATCOM delay} represents the one-way delay of the SATCOM connection, where a static value, or a list of values stating the change over time, can be configured.
Moreover, \emph{loss} configures the loss of the SATCOM connection, and \emph{attenuation} the signal damping.

\textbf{Transport parameters.}
The transport parameters are specific to the measurement components \emph{Client} and \emph{Server}, as well as the PEPs \emph{Proxy-ST} and \emph{Proxy-GW}.
Hence, every parameter can be individually specified for each transport component, which enables the optimization of the SATCOM connection using \emph{Proxy-ST} and \emph{Proxy-GW} independently of the \emph{Client} and \emph{Server} parametrization.
The \emph{Congestion Control} parameter configures the \ac{cca}, where \texttt{Cubic} and \texttt{NewReno} are the available options.
While \texttt{Cubic} is the default \ac{cca} in most of today's operating systems, \texttt{NewReno} is preferred on high RTT connections as experienced in SATCOM networks due to its less aggressive \emph{congestion window} growth~\cite{rfc8312}.
However, more optimized \ac{cca} implementations (e.g., \texttt{Hybla, BBR}) exist for satellite networks~\cite{sat.cc}.
Since the QUIC implementation we use is currently limited to \texttt{Cubic} and \texttt{NewReno} (see §~~\ref{sub:quic-pep} and §~\ref{sub:qperf}), we restrict the available options in order to achieve comparable results between QUIC and TCP (see §~\ref{sec:limitations}).
Moreover, the \emph{\ac{iw}} parameter configures the initial window, while the \emph{ACK Frequency} parameter is specific to QUIC and enables support for the \emph{QUIC Acknowledgement Frequency} extension~\cite{ietf-quic-ack-frequency}.

\begin{table}[t]
    \caption{Emulation parameters for \emph{scenario} configurations}
    \definecolor{rowgray}{gray}{0.9}
    \centering
    \begin{tabular}{lll}
        \toprule
            \textbf{Category} & \textbf{Parameter} & \textbf{Values}\\
        \midrule
            Testbed & Internet delay & \emph{ms}, static \\
                & SATCOM delay & \emph{ms}, static or dynamic \\
                & SATCOM loss & \emph{percentage} \\
                & SATCOM attenuation & \emph{db} \\
            \midrule
            Transport & Congestion Control & \emph{CUBIC}, \emph{NewReno} \\
                & Initial Window & \emph{maximum packets} \\
                & ACK Frequency (QUIC) & \emph{ack-freq parameters}\\
        \bottomrule
    \end{tabular}
    \label{tab:scenario-parameters}
    \vspace{-0.5em}
\end{table}

\subsection{Measurement Types}
\label{sub:measurement-types}
Within each \emph{scenario}, multiple measurements are performed using QUIC, TCP, HTTP/3 (which uses QUIC), and HTTP/1.1 (which uses TCP), each with and without the aid of a \ac{pep}.
We acknowledge that the number of websites supporting HTTP/2 is rising~\cite{h2usage}, and we will additionally evaluate HTTP/2 in a future study (see §~\ref{sec:limitations}).

\textbf{QUIC.}
For the \emph{QUIC} measurement type, we developed the \emph{QUIC performance measurement tool} (see §~\ref{sub:qperf}) which measures the \emph{connection establishment time}, the \emph{time to first byte}, the \emph{congestion window}, and 
the \emph{goodput} with and without the usage of the \emph{QUIC PEP} (see §~\ref{sub:quic-pep}).

\textbf{TCP.}
The \emph{TCP} measurement type uses \emph{iperf3} to measure the \emph{congestion window} on the server-side, and the \emph{goodput} on the client-side, optionally proxied by the open-source TCP \ac{pep} \emph{PEPsal}~\cite{pepsal_paper}.
Moreover, the \emph{connection establishment time} and the \emph{time to first byte} are measured using \emph{curl} on the client connecting to a \emph{nginx} web server on the server.

\textbf{HTTP/3.}
The \emph{HTTP/3} measurement type leverages the \emph{H2O} web server.
While \emph{H2O} serves an arbitrary website, it is accessed using \emph{HTTP/3} by a \emph{Chromium} web browser controlled by \emph{Selenium} on the client-side, with or without using the \emph{QUIC PEP} (see §~\ref{sub:quic-pep}).
Using the \emph{Performance Navigation Timing} API~\cite{performance-navigation-timing}, various \emph{web performance metrics} like \emph{Response Start}, \emph{First Contentful Paint}, and \emph{Page Load Time}, are measured.

\textbf{HTTP/1.1.}
For \emph{HTTP/1.1}, we measure the same web performance metrics using the same tools as with \emph{HTTP/3}, optionally proxied by the \emph{PEPsal} TCP PEP.

\subsection{QUIC PEP}
\label{sub:quic-pep}
With TCP headers unencrypted, TCP \acp{pep} can be deployed in \ac{satcom} networks for transport layer connection optimizations based on connection splitting (see §~\ref{sec:introduction}).
With its mandatory header encryption, connection splitting is not applicable to QUIC in a similar fashion and, thus, QUIC connections are inherently end-to-end.
To evaluate if QUIC over \ac{satcom} networks could benefit from transport layer optimizations as performed by TCP \acp{pep}, we develop a proxy to enable connection splitting of QUIC connections: The proxy receives incoming connections, establishes a new connection to a predefined destination, and forwards data between both connections while directly mapping the stream IDs.
This simplistic design breaks end-to-end encryption and gives the proxy access to the decrypted user data.
We therefore explicitly note that the implementation of \emph{QUIC PEP} is a proof-of-concept to evaluate if QUIC over \ac{satcom} networks could benefit from transport layer optimizations; see §~\ref{sec:limitations} for a detailed discussion.
This design facilitates concatenating multiple proxies and decouples the congestion control loops of different path segments; hence, transport layer connections can be optimized for their respective path segment properties.

As a basis for \emph{QUIC PEP}, the \emph{quicly}~\cite{quicly} implementation is used.
In the \emph{default} operation mode, which is used throughout the paper until otherwise noted, the \emph{QUIC PEP} performs the handshakes of incoming connections in parallel with the connection establishment with the next hop, which can be an upstream proxy, or a target server.
Using this design, the connection establishment is parallelized, and the client is able to send data before the connection to the target server is established; hence, the time until the server, and subsequently the client, receive the data, is improved.
To enable the proxying of HTTP/3 connections, the \emph{h3-capable} operation mode disables this parallelization, and requires a handshake with the next hop to be completed before completing the handshake with the previous hop.
This requirement traces back to the server-initiated unidirectional HTTP/3 \emph{control} as well as the \emph{header compression} encode/decode streams, which must be received by the client before the HTTP/3 request is sent in order to avoid connection failure caused by state mismatch between server and client.

For transport layer optimization, \emph{QUIC PEP} offers the parametrization of the \emph{\ac{cca}}, the \emph{\ac{iw}}, and the \emph{QUIC version}, and also supports the \emph{QUIC Acknowledgement Frequency} extension~\cite{ietf-quic-ack-frequency}.

\subsection{QUIC Performance Measurement Tool}
\label{sub:qperf}
In order to evaluate QUIC performance, we develop a \emph{QUIC performance measurement tool} consisting of a client and a server module.
Using a client-initiated connection, the \emph{connection establishment time} as well as the \emph{time to first byte} between client and server is measured.
Subsequently, the server sends arbitrary data back to the client, where the corresponding \emph{congestion window} is evaluated on the server-side, and the resulting \emph{goodput} on the client-side.
The measurement tool is also based on \emph{quicly} and offers parametrization options similar to \emph{QUIC PEP}: the \emph{\ac{cca}}, the \emph{\ac{iw}}, and the \emph{QUIC version}.
Moreover, it offers support for the \emph{QUIC Acknowledgement Frequency} extension~\cite{ietf-quic-ack-frequency}, and incorporates the \emph{qlog}~\cite{qlog} logging schema to facilitate measurement analysis.

\subsection{Validation}
\label{sub:validation}

We evaluate the SATCOM emulation testbed in order to validate its \emph{functionality}, \emph{accuracy}, and \emph{reproducibility}.
We perform a functional validation by running the SATCOM emulation testbed with different \emph{scenario} configurations for every \emph{measurement type}, where each \emph{scenario} is run 100 times.
Before each \emph{scenario} is run, we execute \emph{ICMP} control measurements, and capture the packets on both the client and server side using \emph{tcpdump} during the emulation.
Evaluating the \emph{ICMP} measurements as well as the packet captures, we find that the \emph{loss} and \emph{delay} characteristics are accurately emulated.
However, we observe an increased SATCOM \emph{delay} for \emph{connection establishment time} and \emph{time to first byte} of $\sim$10ms for TCP-based connections and $\sim$25ms for QUIC-based connections.
While we are not able to attribute this observation to a single cause, an analysis identified that the delay is added by the OpenSAND SATCOM emulation.
Following the successful completion of the emulation runs, we evaluate the \emph{accuracy} via the automated visualization of all \emph{measurement types}, augmented by a manual analysis of the packet captures, the client, server, and proxy logs, and the CPU and RAM utilization metadata.
We find, that the emulation is not limited by neither CPU nor RAM utilization, and that the \emph{goodput} converges to the maximum link-layer goodput as configured by the SATCOM components.
Moreover, we validate the \emph{reproducibility} of the SATCOM emulation testbed by comparing the results of 3 different Linux systems, finding identical results for all \emph{scenarios} and \emph{measurement types}.


\section{Evaluation}
\label{sec:quic-proxies}

With a validated emulation testbed in place, we now proceed to evaluate if QUIC benefits from transport layer optimizations through \acp{pep}, analyzing \emph{goodput} as well as \emph{web performance} characteristics over multiple SATCOM network configurations.
The emulation is run on an Ubuntu 18.04 system with Kernel 5.4.0, featuring 2 Intel Xeon E5-2643 6-Core CPUs and 128GB of RAM.
The SATCOM components are configured with a clear-sky Signal-to-Noise Ratio of 20dB, a constant QPSK 1/4 modulation, a roll-off factor of 0.25, a return-band (client to server direction) bandwidth of 20MHz, and a forward-band (server to client direction) bandwidth of 50MHz, resulting in a maximum forward-band link-layer goodput of 20Mbps.
Using the validation (see §~\ref{sub:validation}), we ensure that the emulation is not influenced by neither the hardware nor the configuration. 


\begin{figure*}[t]
	\centering
	\includegraphics[width=\linewidth]{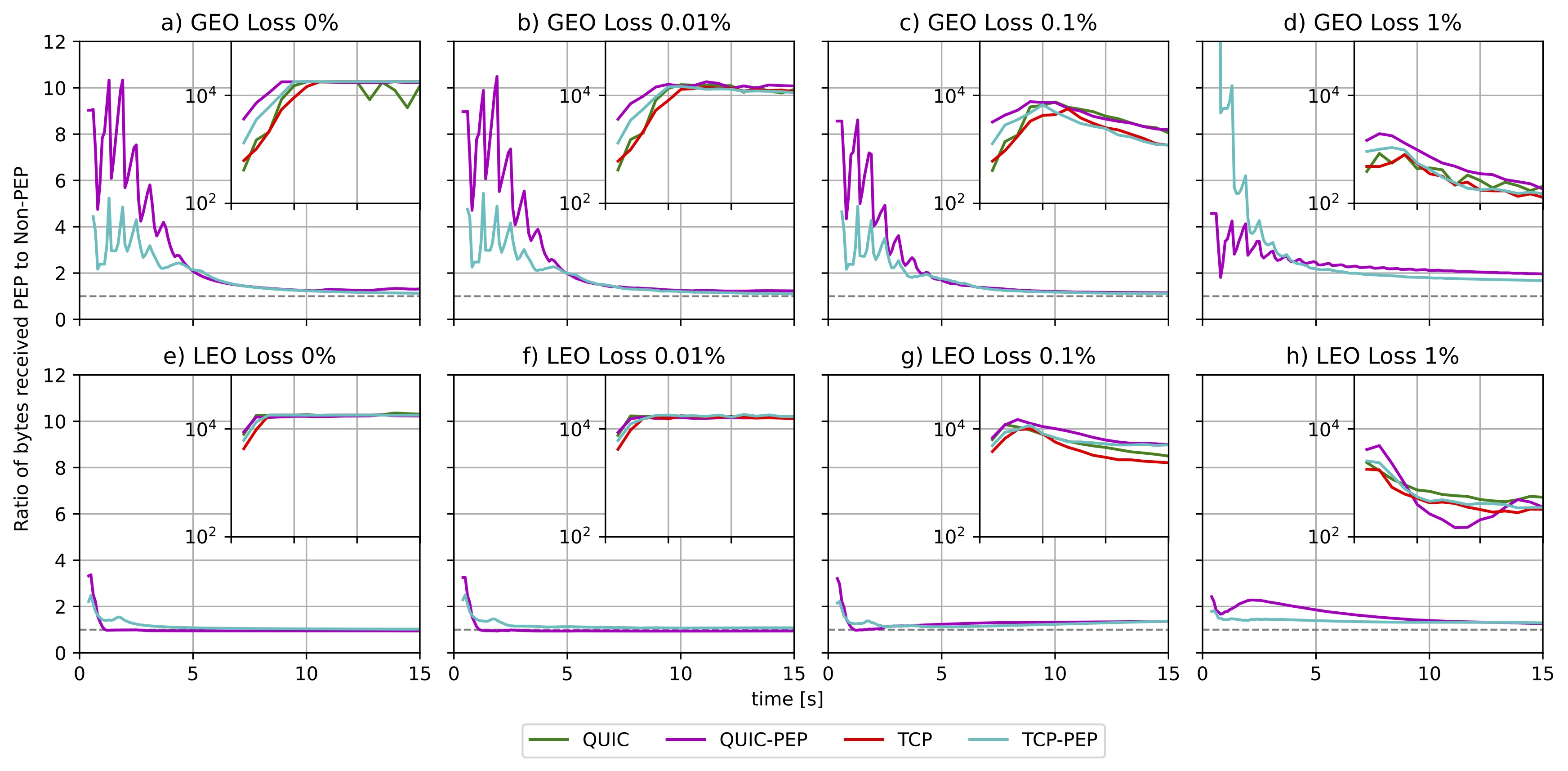}
	\vspace{-2em}
	\caption{Ratio of bytes received PEP to Non-PEP (main plots) and absolute goodput (embedded plots) over 15s for GEO (top row) as well as LEO (bottom row) satellite orbits for real world (0, 0.01\%) and edge case (0.1, 1\%) loss conditions (columns from left to right).
	A factor of 1 represents no improvement, a factor of >1 shows a benefit by the usage of a PEP, and a factor of <1 represents a degradation of PEP in comparison to Non-PEP connections.
	} 
	\label{fig:goodput}
	\vspace{-1em}
\end{figure*}

Using the \emph{scenario} model (see §~\ref{sub:scenarios}), we differentiate between the two satellite orbits GEO and LEO, where we set the \emph{SATCOM} one-way delay to 250ms for GEO as derived from the speed of light in a vacuum.
In order to determine a typical one-way delay for LEO orbits, we perform $\sim$600k RTT measurements from a vantage point in Central Europe over a period of one week using Starlink~\cite{starlink}, where we find a median first hop RTT of $\sim$32ms (mean $\sim$33ms).
While we acknowledge that the first hop RTT in LEO constellations changes due to the movement of the satellites, we observe relatively constant RTTs for at least 30 consecutive seconds, which is in line with the observations of Kassing et al.~\cite{hypatia} and Pavur et al.~\cite{qpep}.
As our measurements do not exceed a duration of 20s, we set the \emph{SATCOM} LEO one-way delay to a static value of 16ms (1/2 RTT).
In addition to the \emph{SATCOM} delay, we set the \emph{Internet} one-way delay to 40ms for both GEO and LEO orbits in order to emulate the terrestrial distance between the satellite ground station and the target server.
Hence, the delay configuration results in a GEO RTT of 580ms and a LEO RTT of 112ms.
%
Lastly, the \emph{attenuation} is configured with 0dB, i.e., no signal damping, and the packet \emph{loss} rate is chosen from 0, 0.01, 0.1, and 1\%.
While the transport layer loss is considered to range from nearly 0\% (where almost all errors are corrected by the link-layer) up to 0.01\% in real world satellite conditions~\cite{qpep}, we include 0.1 and 1\% loss conditions in order to evaluate edge cases.

We further differentiate between the Non-PEP and PEP \emph{transport} parameter configurations.
For Non-PEP, we configure both client and server with default values used in both QUIC and TCP stacks, where the \emph{\ac{cca}} is set to \texttt{Cubic} and the \emph{\ac{iw}} to 10.
For PEP, we use the identical settings as for Non-Pep for client and server, but optimize the SATCOM connection using PEPs by setting the \emph{\ac{cca}} to \texttt{NewReno} and the \emph{\ac{iw}} to 100 for both QUIC and TCP.
Finally, we use QUIC version 1 for all QUIC configurations.

With the above \emph{scenarios}, we emulate a typical \ac{satcom} use-case with varying orbits as well as loss characteristics, enabling a direct comparison between PEP and Non-PEP connections using QUIC, TCP, HTTP/3, and HTTP/1.1.
While the emulation testbed also offers more advanced \emph{scenario} configurations (see §~\ref{sec:satcom-measurement-testbed}), we limit our evaluation to the configurations presented above to get a deeper understanding of the \emph{\ac{cca}} and \emph{\ac{iw}} optimization potential, which are traditionally optimized by TCP-PEPs~\cite{rfc3135}.

\subsection{Goodput}
\label{sub:goodput}

We evaluate the \emph{goodput} over time presented in Fig.~\ref{fig:goodput} by analyzing the ratio of bytes received between PEP and Non-PEP (main plots) and absolute goodput (embedded plots) for GEO (top row) as well as LEO (bottom row) satellite orbits, each with 0, 0.01, 0.1, and 1\% loss (columns from left to right).
With a duration of 15s, a typical file-download use-case is evaluated.
The measurements are repeated 100 times and we present the averages over all measurement runs.
The main plots show the relative difference of bytes received over time between PEP and Non-PEP connections.
A factor of 1 (dashed line) represents no improvement, a factor of >1 shows a benefit by the usage of a PEP, and a factor of <1 represents a degradation of PEP in comparison to Non-PEP connections.
Moreover, the embedded plots show the absolute goodput over time for QUIC, QUIC-PEP, TCP, and TCP-PEP connections.

Evaluating the GEO measurements (Fig.~\ref{fig:goodput} top row, \emph{a} to \emph{d}), we observe a benefit of PEPs over all loss configurations, where the factor of bytes received increases up to 10$\times$ for PEP QUIC connections (magenta line).
The benefit is more pronounced in the first 5 seconds; we attribute this improvement to the \emph{\ac{iw}} optimization of the PEPs.
On high RTT connections as experienced in SATCOM networks, the slow start can become ACK-locked: While the bytes in flight are limited by the \emph{congestion window}, the sender has to pause the transmission until an ACK is received.
When we increase the \emph{\ac{iw}} by 10$\times$, more data can initially be sent before the receipt of an ACK increases the \emph{congestion window}, resulting in a faster slow start until the slow start threshold is reached.
This benefit can also be observed in the embedded plots of the top row showing the absolute goodput over time: The slow start is considerably faster using PEPs (magenta and cyan lines), improving over Non-PEPs (green and red lines) for 0 and 0.01\% loss configurations by up to 2s (Fig.~\ref{fig:goodput} \emph{a} and \emph{b}).
While the edge case loss configurations of 0.1 and 1\% (\emph{c} and \emph{d}) also show benefits from the increased \emph{\ac{iw}} of the PEP connections (magenta and cyan lines), we observe a degradation in goodput after reaching the slow start threshold for both PEP and Non-PEP connections.
The overall goodput is drastically reduced; however, the PEP connections reach the slow start threshold faster.
In addition, we observe a benefit for QUIC connections (green and magenta lines) in comparison to TCP (red and cyan lines), where both QUIC connections result in higher goodput following slow start.
Evaluating 1\% loss (Fig.~\ref{fig:goodput}~\emph{d}) for TCP connections, we observe an increase for TCP-PEP of up to 85$\times$ within the first second of the measurement (cyan line).
While this factor exceeds our increase in \emph{\ac{iw}} of 10$\times$, we attribute this outlier to the \emph{iperf3} measurement tool: While we sample the bytes received in 0.1s intervals, \emph{iperf3} regularly omits the first 1--8 intervals in the 1\% loss configuration, resulting in the observed inaccuracies within the first second.

Evaluating the LEO measurements (Fig.~\ref{fig:goodput} bottom row, \emph{e} to \emph{h}), we again observe a benefit of PEPs over all loss configurations, where the factor of bytes received increases up to 3$\times$ for PEP QUIC connections (magenta line).
This benefit is more pronounced within the first 2 seconds; however, while the increase in \emph{\ac{iw}} results in a faster slow start, the slow start threshold is reached at the same time for PEP (magenta and cyan) and Non-PEP (green and red) connections (bottom row embedded plots).
Moreover, the benefit is less pronounced in comparison to GEO.
Evaluating the 0.1 and 1\% loss edge cases (\emph{g} and \emph{h}), we observe similar trends in comparison to GEO while the overall goodput is drastically reduced.
Furthermore, we see a benefit of PEP connections almost over the complete measurement for 0.1\% loss (Fig.~\ref{fig:goodput} \emph{g}), where QUIC-PEP shows a fast degradation on 1\% loss following slow start (Fig.~\ref{fig:goodput} \emph{h}).

\begin{table*}[t]
    \caption{Median \emph{Response Start (RS)}, \emph{First Contentful Paint (FCP)}, and \emph{Page Load Time (PLT)} in milliseconds of PEP and Non-PEP \\ HTTP/3 (h3) as well as HTTP/1.1 (h1) connection for GEO (top) as well as LEO (bottom) satellite orbits \\ for real world (0, 0.01\%) and edge case (0.1, 1\%) loss conditions (columns from left to right).}
    \resizebox{\linewidth}{!}{%
        \begin{tabular}{ll|rr|rr|rr|rr|rr|rr}
            \toprule
                & {} & \multicolumn{4}{c|}{\textbf{Response Start (RS)}} & \multicolumn{4}{c|}{\textbf{First Contentful Paint (FCP)}} & \multicolumn{4}{c}{\textbf{Page Load Time (PLT)}} \\
            \textbf{Orbit} & \textbf{Protocol} / \textbf{Loss}  &     \textbf{0\%} &  \textbf{0.01\%} &  \textbf{0.1\%} &  \textbf{1\%} &     \textbf{0\%} &  \textbf{0.01\%} &  \textbf{0.1\%} &  \textbf{1\%} &     \textbf{0\%} &  \textbf{0.01\%} &  \textbf{0.1\%} &  \textbf{1\%} \\
            \midrule
            GEO & h3&     1314 &  1310 &  1312 &  1321 &    2759 &  2760 &  2773 &  2861 &    7669 &   7671 &   7739 &  16865 \\
                &\cellcolor[gray]{0.9}h3-PEP &\cellcolor[gray]{0.9}     2717 &\cellcolor[gray]{0.9}  2715 &\cellcolor[gray]{0.9}  2717 &\cellcolor[gray]{0.9}  2727 &\cellcolor[gray]{0.9}    3581 &\cellcolor[gray]{0.9}  3575 &\cellcolor[gray]{0.9}  3581 &\cellcolor[gray]{0.9}  4196 &\cellcolor[gray]{0.9}    7344 &\cellcolor[gray]{0.9}   7336 &\cellcolor[gray]{0.9}   7522 &\cellcolor[gray]{0.9}   9409 \\
                & h1 &     1222 &  1220 &  1214 &  1221 &    3841 &  3844 &  3832 &  4172 &   10806 &  10835 &  11176 &  15802 \\
                &\cellcolor[gray]{0.9}h1-PEP &\cellcolor[gray]{0.9}     1215 &\cellcolor[gray]{0.9}  1215 &\cellcolor[gray]{0.9}  1214 &\cellcolor[gray]{0.9}  1215 &\cellcolor[gray]{0.9}    3852 &\cellcolor[gray]{0.9}  3842 &\cellcolor[gray]{0.9}  3836 &\cellcolor[gray]{0.9}  4404 &\cellcolor[gray]{0.9}   11144 &\cellcolor[gray]{0.9}  11124 &\cellcolor[gray]{0.9}  11175 &\cellcolor[gray]{0.9}  13994 \\
            \midrule
            LEO & h3 &      434 &   436 &   435 &   447 &    1408 &  1410 &  1415 &  1489 &    2612 &   2617 &   2649 &   7010 \\
                &\cellcolor[gray]{0.9}h3-PEP &\cellcolor[gray]{0.9}     1418 &\cellcolor[gray]{0.9}  1414 &\cellcolor[gray]{0.9}  1419 &\cellcolor[gray]{0.9}  1416 &\cellcolor[gray]{0.9}    1880 &\cellcolor[gray]{0.9}  1868 &\cellcolor[gray]{0.9}  1882 &\cellcolor[gray]{0.9}  1914 &\cellcolor[gray]{0.9}    3172 &\cellcolor[gray]{0.9}   3173 &\cellcolor[gray]{0.9}   3264 &\cellcolor[gray]{0.9}   3872 \\
                & h1 &      291 &   288 &   294 &   303 &    1138 &  1135 &  1145 &  1329 &    3218 &   3227 &   3302 &   4573 \\
                &\cellcolor[gray]{0.9}h1-pep &\cellcolor[gray]{0.9}      289 &\cellcolor[gray]{0.9}   286 &\cellcolor[gray]{0.9}   287 &\cellcolor[gray]{0.9}   301 &\cellcolor[gray]{0.9}    1095 &\cellcolor[gray]{0.9}  1085 &\cellcolor[gray]{0.9}  1084 &\cellcolor[gray]{0.9}  1356 &\cellcolor[gray]{0.9}    3308 &\cellcolor[gray]{0.9}   3321 &\cellcolor[gray]{0.9}   3367 &\cellcolor[gray]{0.9}   4298 \\
            \bottomrule
        \end{tabular}
    }
    \label{tab:web-performance}
    \vspace{-0.5em}
\end{table*}

\takeaway{
While we observe benefits for PEP connections over all orbits and loss configurations, the improvements are, expectedly, more pronounced on connections with higher RTTs and less loss, where the slow start threshold is reached up to 2s faster in comparison to Non-PEP connections.
The benefits are primarily limited to the slow start phase in real world conditions, which results from the \emph{\ac{iw}} optimization of the PEPs. However, PEP connections can also lead to a degradation in goodput on SATCOM connections with shorter RTTs.
}

\subsection{Web Performance}
\label{sub:web-performance}

We evaluate the web performance by analyzing the median values of \emph{Response Start (RS)}, \emph{First Contentful Paint (FCP)}, and \emph{Page Load Time (PLT)} over 100 measurement runs for every orbit, loss, and protocol combination presented in milliseconds in Tab.~\ref{tab:web-performance}.
To enable the proxying of HTTP/3 connections, the \emph{h3-capable} operation mode of \emph{QUIC PEP} is used (see §~\ref{sub:quic-pep}); hence, a handshake with the next hop has to be completed before completing the handshake with the previous hop, resulting in a sequential connection establishment.
Moreover, we use HTTP/1.1 without encryption in contrast to the TLS 1.3 encrypted HTTP/3.
While the overhead of the TLS encryption adds 1 (in case of TLS 1.3), respective 2 (in case of TLS 1.2) RTTs to the TCP connection establishment of HTTP/1.1, the overhead is systematic. 
Hence, we leverage unencrypted HTTP/1.1 as a performance oriented baseline for our comparison to HTTP/3, where our results can be extrapolated for TLS 1.3 / TLS 1.2 encrypted HTTP/1.1 by adding 1, respective 2, RTTs.

For our analysis, we use the \emph{ETSI Kepler Web Reference Page}~\cite{etsi-kepler} that aims to represent a ``typical'' website.
While we acknowledge that an objective characterization of a typical website is debatable considering the heterogeneity of the web, choosing this ``representative'' website can still serve as a first indication if QUIC is able to benefit from proxies.
A more diverse set of websites will be evaluated in future work (see §~\ref{sec:limitations}).
The default settings of the \emph{Chromium} client and of the \emph{H2O} server are used to transport the website's 75 objects ($\sim$880kb) either over a single QUIC connection with 6 streams in case of HTTP/3 (h3), or over 6 distinct TCP connections in case of HTTP/1.1 (h1). QUIC version 1 and HTTP/3 version draft-29 are used.

First, we take a look at the \emph{Response Start (RS)} web performance metric, which represents the time that passes between the sending of the first packet of the transport handshake and the reception of the first byte of the HTTP response by the client~\cite{web-moz-ttfb}.
Hence, the \emph{RS} is expected to resemble 2 RTTs plus the static one-way overhead added by OpenSAND of $\sim$10ms for TCP and $\sim$25ms for QUIC (see §~\ref{sub:validation}).
Analyzing the GEO \emph{RS} presented in Tab.~\ref{tab:web-performance}, we observe $\sim$1.2s for the TCP-based h1 and h1-PEP protocols, while the QUIC-based h3 is moderately ($\sim$1.3s) and h3-PEP considerably ($\sim$2.7s) slower in comparison.
The duration of \emph{RS} is largely independent of the packet loss rate.
The protocols h1, h1-PEP and h3 show the expected \emph{RS} that roughly equals 2 GEO RTTs of 580ms plus the static overhead added by OpenSAND\@.
The increased \emph{RS} of h3-PEP ($\sim$2.7s) traces back to the \emph{h3-capable} operation mode of \emph{QUIC PEP}:
Because the connections are established sequentially, and multiple message exchanges are required for each connection setup, the h3-PEP \emph{RS} is considerably higher.
Evaluating h3 and h3-PEP, we find additional overheads for QUIC by the processing on the server and the proxies.
Analyzing the \emph{quicly} library used by \emph{QUIC PEP} as well as the \emph{H2O} web server (see §~\ref{sec:satcom-measurement-testbed}), we find that the additional overheads can be attributed to the \texttt{quicly\_accept()} function which accepts new connections, resulting in an increase of 2--6ms per connection establishment.
Looking at the \emph{RS} for the LEO network over all loss configurations, we observe the same trends as in GEO\@: While h1 and h1-PEP are on par with $\sim$290ms, h3 is moderately ($\sim$435ms), and h3-PEP considerably ($\sim$1.4s) slower in comparison.
Our results therefore show that the \emph{RS} of the TCP-based protocols h1 and h1-PEP are faster in comparison to the QUIC-based protocols h3 and h3-PEP\@.

As a second web performance metric we evaluate \emph{First Contentful Paint (FCP)}, which measures the time between the sending of the first packet of the transport handshake and the rendering of the first element by the browser; hence, it represents the user-perceived load speed of a website~\cite{web-moz-fcp}.
Evaluating the GEO \emph{FCP} presented in Tab.~\ref{tab:web-performance}, we observe identical values per protocol over the 0 and 0.01\% loss configurations with  $\sim$2.8s for h3, $\sim$3.6s for h3-PEP, as well as $\sim$3.8s for both h1 and h1-PEP.
While the QUIC-based protocols h3 and h3-PEP show a slower \emph{RS}, both improve over their TCP counterparts for \emph{FCP}.
We attribute this improvement to QUIC's multiplexing capabilities\@: While 6 distinct TCP connections are established by \emph{Chromium} for h1 and h1-PEP to avoid head-of-line blocking, 6 streams are sent over a single QUIC connection for h3 and h3-PEP\@.
Hence, the overhead of connection establishment is considerably reduced, leading to faster \emph{FCP} for h3 and h3-PEP in comparison to h1 and h1-PEP\@.
Moreover, we also observe a relative improvement comparing \emph{RS} and \emph{FCP} for h3 and h3-PEP\@: While the overhead of the sequential connection establishment of h3-PEP results in a \emph{RS} slowdown of $\sim$1.4s, this difference is reduced to $\sim$820ms for \emph{FCP}.
Looking at GEO 0.1\% loss, we observe identical values in comparison to 0 and 0.01\%, while 1\% loss shows an increase of up to $\sim$600ms for h3-PEP as well as h1-PEP\@.
We observe similar trends in the \emph{FCP} of GEO and LEO orbits for 0 and 0.01\% loss.
However, due to the reduced RTT in the LEO orbit, h3-PEP can not yet overcome the initial overhead of the sequential connection establishment in comparison to h1 and h1-PEP\@.
With a \emph{FCP} of $\sim$1.9s, h3-PEP falls short of h1 and h1-PEP ($\sim$1.1s).
Moreover, h3-PEP also falls short of h3 by $\sim$470ms, but still improves over their \emph{RS} difference of $\sim$1s.
Considering LEO 0.1\% loss, we observe slightly increased values in comparison to 0 and 0.01\%, where 1\% loss shows an increase of up to $\sim$260ms for h1-PEP\@.

Our third web performance metric is the \emph{Page Load Time (PLT)} (see Tab.~\ref{tab:web-performance}), which represents the time between the sending of the first packet of the transport handshake until all content of the website is received by the browser~\cite{web-moz-plt}.
Looking at GEO's 0 and 0.01\% loss configurations, we observe the fastest \emph{PLT} for h3-PEP with $\sim$7.3s, followed by h3 ($\sim$7.7s), h1 ($\sim$10.8s), and h1-PEP ($\sim$11.1s).
We again attribute the performance benefit of the QUIC-based h3 and h3-PEP connections to QUIC's stream multiplexing feature.
Moreover, we observe the break-even of the initial overhead of the sequential connection establishment for h3-PEP, outperforming h3 by $\sim$330ms, and h1 as well as h1-PEP by more than 3s; hence, we see a benefit of PEPs for QUIC connections in GEO orbits.
At higher loss rates, the h3-PEP performance advantage increases: With 1\% loss, h3-PEP finishes loading more than 7s earlier than h3, and $\sim$4.5s earlier than h1-PEP\@.
Comparing GEO h1 and h1-PEP, we observe that h1 is slightly faster for 0 and 0.01\% loss configurations, while they are on par for 0.1\%; however, h1-PEP shows an improvement of $\sim$2s over h1 for 1\% loss.
Evaluating \emph{PLT} for LEO orbits for 0 and 0.01\% loss, we observe similar trends in comparison to GEO.
While h1 ($\sim$3220ms) and h1-PEP ($\sim$3315ms) show the slowest \emph{PLT}, h3-PEP is faster with $\sim$3170ms, only outperformed by h3 with $\sim$2615ms.
Considering the reduced RTT in LEO orbits, h3-PEP is not able to overcome the initial overhead of the sequential connection establishment in comparison to h3, but improves on h1 and h1-PEP by $\sim$50ms, respective $\sim$145ms.
Looking at 0.1\% loss, this improvement becomes more distinct, where on 1\% loss h3-PEP outperforms h1 and h1-PEP by more than 400ms, and even improves over h3 by more than 3s.

\takeaway{
Due to QUIC's multiplexing capabilities, we observe an improvement of the QUIC-based protocols h3 and h3-PEP over both h1 and h1-PEP, where h3 and h3-PEP achieve faster \emph{FCP} and \emph{PLT} in GEO orbits, as well as faster \emph{PLT} in LEO orbits.
Moreover, while the initial overhead of the sequential connection establishment of h3-PEP leads to slower \emph{RS} and \emph{FCP} in comparison to h3, h3-PEP improves over h3 for \emph{PLT} in GEO orbits: With a reduction of $\sim$330ms for real world conditions, and over 7s in edge cases, we observe a benefit of PEPs for QUIC connections.
}

\section{Limitations and Future Work}
\label{sec:limitations}

While our \emph{QUIC PEP} realizes transport layer optimizations by means of connection splitting, the implementation is considered a proof-of-concept as the proxy is able to access the decrypted data, yet sufficient for the purpose of exploring possible benefits of QUIC PEPs.
Hence, we are currently exploring schemes where the QUIC payload remains end-to-end encrypted and only selected control information are exposed to the proxies.
Traffic tunneling mechanisms as discussed in the IETF MASQUE WG \cite{ietf-masque-connect-udp} preserve encryption and allow for an independent congestion control loop between these PEPs, but still run nested congestion control end-to-end, not providing connection splitting.
Yet, the basic signaling mechanisms provide a well-defined means for interacting with proxies and thus could be leveraged for extensions towards QUIC PEPs, with further mechanisms for preserving end-to-end encryption to be explored.

Moreover, the PEPed QUIC connections are currently established per client request, offering additional optimization potential: By leveraging \emph{0-RTT} between the proxies, the time required for the connection establishment can be further reduced.
Considering the \acp{cca}, both \emph{QUIC PEP} and the \emph{QUIC performance measurement tool} currently only offer \texttt{Cubic} and \texttt{NewReno}.
While more optimized implementations exist for satellite networks (e.g., \texttt{Hybla, BBR}), we will integrate and evaluate additional \emph{\acp{cca}} in a future study.
We also plan to incorporate the \emph{Acknowledgement Frequency} extension~\cite{ietf-quic-ack-frequency} in order to reduce the number of acknowledgements sent, as well as the QUIC \emph{BDP Frame} extension~\cite{kuhn2021} to accelerate the goodput ramp-up on repeated connections; both show promising results for SATCOM networks.

While our findings revealed that QUIC connections are able to benefit from proxies through transport layer optimization for both \emph{goodput} and \emph{web performance}, we acknowledge that our findings on \emph{web performance} are (naturally) influenced by the website selection.
Therefore, we seek to evaluate a more diverse set of websites in a future study in order to generalize our findings, where we will also incorporate HTTP/2.

\section{Related Work}
\label{sec:related-work}


%


Several papers~\cite{quic-perf-sat-iwcmc,zhang2019,gquic_satcom,satellite-quic,kuhn2020,quic.acks.satcom} investigate the usage of QUIC in satellite networks.
They find that QUIC realizes lower Page Load Times on the web compared to TCP, mainly due to its faster connection setup~\cite{quic-perf-sat-iwcmc,zhang2019}.
Also, the utility of PEPs for improving the performance of TCP over long-delay satellite links has long been known~\cite{rfc3449}.
The studies that compare the performance of TCP using PEPs (TCP-PEP) with QUIC, find that TCP-PEP generally outperforms QUIC for larger transfers~\cite{gquic_satcom,satellite-quic,kuhn2020,quic.acks.satcom}.

Some authors argue that specific tuning, such as an increase of the initial window, can improve QUIC's performance~\cite{gquic_satcom,kuhn2020,quic.acks.satcom}.
Custura et al.~\cite{quic.acks.satcom} investigate how decreasing the acknowledgement frequency can reduce the control overhead and in some cases improve performance.
Kuhn et al.~\cite{kuhn2021} introduce the QUIC \emph{BDP Frame} extension that accelerates the throughput ramp-up on repeated connections over long-delay satellite links.

We take a different approach to examine the impact of transport-layer optimizations on the performance of QUIC over GEO and LEO links by developing a proof-of-concept QUIC-PEP\@.
Previous work on intermediaries with QUIC for SATCOM is sparse
and the closest related work is from Pavur et al.~\cite{qpep} on QPEP, who multiplex TCP connections over a single QUIC connection between ground terminals and thereby significantly reduce Page Load Times.
They, however, study a long-standing QUIC connection and therefore exclude the connection setup.
While \texttt{netem} is commonly applied at a single point to model a satellite link \cite{quic-perf-sat-iwcmc,quic_sat_bbr,zhang2019},
we use OpenSAND~\cite{opensand} in conjunction with netem and induce delay at different points along the network path to model our emulation closer to reality.

A large fraction of previous studies~\cite{long-look-at-quic,http-over-quic,when-quic-meets-tcp,cook2017} examines the performance of Google's QUIC flavor (gQUIC) that differs in fundamental points, such as the cryptographic handshake, from the IETF specification of QUIC~\cite{nottingham2018}.
%
Recent web performance evaluations show that HTTP/3 using IETF QUIC does not necessarily perform better than HTTP/2, which builds upon TCP~\cite{saif2021,yu2021}.
Saif et al.~\cite{saif2021} find that TCP performs better except if loss is present. In this case, QUIC's design shows its strength by reducing the impact of head-of-line blocking.
Similarly, Yu et al.~\cite{yu2021} present mixed findings and highlight the impact of configuration choices on QUIC's performance.
We add a SATCOM perspective to the research space of HTTP/3 performance by analyzing common web performance metrics such as the Page Load Time.



\section{Conclusion}
\label{sec:conclusion}

In this paper, we presented a satellite emulation testbed which enables reproducible QUIC, TCP, HTTP/3, and HTTP/1.1 measurements by using our specifically designed QUIC PEP as well as QUIC performance measurement implementations.
Using the emulation testbed, we carried out an extensive emulation study for LEO and GEO satellites, exploring various combinations of link characteristics.
We found, that the slow start threshold is reached up to 2s faster for QUIC PEP in comparison to QUIC Non-PEP, where the improvements are more pronounced on connections with higher RTTs.
Moreover, we showed that HTTP/3 and HTTP/3-PEP outperform HTTP/1.1 and HTTP/1.1-PEP in multiple \textit{web performance} scenarios which we attribute to QUIC's multiplexing capabilities.
In addition, HTTP/3-PEP also improves over HTTP/3 for \emph{Page Load Time} in GEO orbits by over 7s in edge cases.
Hence, our findings show that PEPs can be beneficial for QUIC connections and warrant further exploration:
While the presented QUIC PEP is considered a proof-of-concept, the basic signaling mechanisms discussed in the IETF MASQUE WG provide well-defined means for interacting with proxies and thus could be leveraged for extensions towards end-to-end encrypted QUIC PEPs.

\section*{Acknowledgements}
We thank Robert Brüning, Felix Beil, and Curt Polack for their valuable efforts, as well as the anonymous reviewers for their insightful feedback.


\bibliographystyle{IEEEtran}
\bibliography{index}

\begin{thebibliography}{10}
\providecommand{\url}[1]{#1}
\csname url@samestyle\endcsname
\providecommand{\newblock}{\relax}
\providecommand{\bibinfo}[2]{#2}
\providecommand{\BIBentrySTDinterwordspacing}{\spaceskip=0pt\relax}
\providecommand{\BIBentryALTinterwordstretchfactor}{4}
\providecommand{\BIBentryALTinterwordspacing}{\spaceskip=\fontdimen2\font plus
\BIBentryALTinterwordstretchfactor\fontdimen3\font minus \fontdimen4\font\relax}
\providecommand{\BIBforeignlanguage}[2]{{%
\expandafter\ifx\csname l@#1\endcsname\relax
\typeout{** WARNING: IEEEtran.bst: No hyphenation pattern has been}%
\typeout{** loaded for the language `#1'. Using the pattern for}%
\typeout{** the default language instead.}%
\else
\language=\csname l@#1\endcsname
\fi
#2}}
\providecommand{\BIBdecl}{\relax}
\BIBdecl

\bibitem{starlink}
\BIBentryALTinterwordspacing
``{Starlink},'' [Accessed 2022-Apr-30]. [Online]. Available: \url{https://www.starlink.com}
\BIBentrySTDinterwordspacing

\bibitem{kuiper}
\BIBentryALTinterwordspacing
A.~Boyle, ``Amazon to offer broadband access from orbit with 3,236-satellite `project kuiper' constellation,'' [Accessed 2022-Apr-30]. [Online]. Available: \url{https://www.geekwire.com/2019/amazon-project-kuiper-broadband-satellite/}
\BIBentrySTDinterwordspacing

\bibitem{oneweb}
\BIBentryALTinterwordspacing
OneWeb, ``Oneweb: Connect with ease,'' [Accessed 2022-Apr-30]. [Online]. Available: \url{https://oneweb.net/connect_with_ease}
\BIBentrySTDinterwordspacing

\bibitem{telesat}
\BIBentryALTinterwordspacing
Telesat, ``Telesat: Global satellite operators,'' [Accessed 2022-Apr-30]. [Online]. Available: \url{https://www.telesat.com/}
\BIBentrySTDinterwordspacing

\bibitem{iridium}
\BIBentryALTinterwordspacing
Iridium, ``Staying connected,'' [Accessed 2022-Apr-30]. [Online]. Available: \url{https://www.iridium.com}
\BIBentrySTDinterwordspacing

\bibitem{rfc3135}
\BIBentryALTinterwordspacing
J.~Griner \emph{et~al.}, ``{Performance Enhancing Proxies Intended to Mitigate Link-Related Degradations},'' RFC 3135, Jun. 2001. [Online]. Available: \url{https://www.rfc-editor.org/info/rfc3135}
\BIBentrySTDinterwordspacing

\bibitem{rfc9000}
\BIBentryALTinterwordspacing
J.~Iyengar and M.~Thomson, ``{QUIC: A UDP-Based Multiplexed and Secure Transport},'' RFC 9000, May 2021. [Online]. Available: \url{https://www.rfc-editor.org/info/rfc9000}
\BIBentrySTDinterwordspacing

\bibitem{rfc9001}
\BIBentryALTinterwordspacing
M.~Thomson and S.~Turner, ``{Using TLS to Secure QUIC},'' RFC 9001, May 2021. [Online]. Available: \url{https://www.rfc-editor.org/info/rfc9001}
\BIBentrySTDinterwordspacing

\bibitem{rfc9002}
\BIBentryALTinterwordspacing
J.~Iyengar and I.~Swett, ``{QUIC Loss Detection and Congestion Control},'' RFC 9002, May 2021. [Online]. Available: \url{https://www.rfc-editor.org/info/rfc9002}
\BIBentrySTDinterwordspacing

\bibitem{masque}
\BIBentryALTinterwordspacing
``{Multiplexed Application Substrate over QUIC Encryption (masque)},'' [Accessed 2022-Apr-30]. [Online]. Available: \url{https://datatracker.ietf.org/wg/masque/about/}
\BIBentrySTDinterwordspacing

\bibitem{opensand}
\BIBentryALTinterwordspacing
``{OpenSAND SATCOM emulation},'' [Accessed 2022-Apr-30]. [Online]. Available: \url{https://opensand.org/}
\BIBentrySTDinterwordspacing

\bibitem{dvbs2}
\BIBentryALTinterwordspacing
``{DVB-S2 specifications},'' [Accessed 2022-Apr-30]. [Online]. Available: \url{https://www.etsi.org/technologies/dvb-s-s2}
\BIBentrySTDinterwordspacing

\bibitem{trustable_sat_emulations}
\BIBentryALTinterwordspacing
A.~Auger \emph{et~al.}, ``{Making Trustable Satellite Experiments: An Application to a VoIP Scenario},'' in \emph{{VTC}}, 2019-Spring. [Online]. Available: \url{https://doi.org/10.1109/VTCSpring.2019.8746404}
\BIBentrySTDinterwordspacing

\bibitem{qpep}
\BIBentryALTinterwordspacing
J.~Pavur \emph{et~al.}, ``{QPEP: An Actionable Approach to Secure and Performant Broadband From Geostationary Orbit},'' in \emph{{NDSS}}, 2021. [Online]. Available: \url{https://doi.org/10.14722/ndss.2021.24074}
\BIBentrySTDinterwordspacing

\bibitem{opensand_ref_1}
\BIBentryALTinterwordspacing
C.~Baudoin and F.~Arnal, ``{Overview of Platine emulation testbed and its utilization to support DVB-RCS/S2 evolutions},'' \emph{ASMS}, 2010. [Online]. Available: \url{https://doi.org/10.1109/ASMS-SPSC.2010.5586897}
\BIBentrySTDinterwordspacing

\bibitem{opensand_ref_2}
\BIBentryALTinterwordspacing
F.~Arnal \emph{et~al.}, ``{Handover Management for Hybrid Satellite/Terrestrial Networks},'' \emph{LNICST}, 2013. [Online]. Available: \url{https://doi.org/10.1007/978-3-642-36787-8_12}
\BIBentrySTDinterwordspacing

\bibitem{rfc8312}
\BIBentryALTinterwordspacing
I.~Rhee \emph{et~al.}, ``{CUBIC for Fast Long-Distance Networks},'' RFC 8312, Feb. 2018. [Online]. Available: \url{https://doi.org/10.17487/RFC8312}
\BIBentrySTDinterwordspacing

\bibitem{sat.cc}
\BIBentryALTinterwordspacing
S.~o. Claypool, ``{Comparison of TCP Congestion Control Performance over a Satellite Network},'' \emph{PAM}, 2021. [Online]. Available: \url{https://doi.org/10.1007/978-3-030-72582-2_29}
\BIBentrySTDinterwordspacing

\bibitem{ietf-quic-ack-frequency}
\BIBentryALTinterwordspacing
J.~Iyengar and I.~Swett, ``{QUIC Acknowledgement Frequency},'' IETF, Internet-Draft draft-ietf-quic-ack-frequency-01, Oct. 2021, work in Progress. [Online]. Available: \url{https://datatracker.ietf.org/doc/draft-ietf-quic-ack-frequency/}
\BIBentrySTDinterwordspacing

\bibitem{h2usage}
\BIBentryALTinterwordspacing
``{HTTP usage statistics},'' [Accessed 2022-Apr-30]. [Online]. Available: \url{https://w3techs.com/technologies/history_overview/site_element/all/y}
\BIBentrySTDinterwordspacing

\bibitem{pepsal_paper}
\BIBentryALTinterwordspacing
C.~Caini \emph{et~al.}, ``{PEPsal: a Performance Enhancing Proxy designed for TCP satellite connections},'' \emph{VTC}, 2006. [Online]. Available: \url{https://doi.org/10.1109/VETECS.2006.1683339}
\BIBentrySTDinterwordspacing

\bibitem{performance-navigation-timing}
\BIBentryALTinterwordspacing
``{PerformanceNavigationTiming API},'' [Accessed 2022-Apr-30]. [Online]. Available: \url{https://w3c.github.io/navigation-timing/#process}
\BIBentrySTDinterwordspacing

\bibitem{quicly}
\BIBentryALTinterwordspacing
``{Quicly QUIC implementation},'' [Accessed 2022-Apr-30]. [Online]. Available: \url{https://github.com/h2o/quicly}
\BIBentrySTDinterwordspacing

\bibitem{qlog}
\BIBentryALTinterwordspacing
R.~Marx \emph{et~al.}, ``{Debugging QUIC and HTTP/3 with Qlog and Qvis},'' \emph{ANRW}, 2020. [Online]. Available: \url{https://doi.org/10.1145/3404868.3406663}
\BIBentrySTDinterwordspacing

\bibitem{hypatia}
\BIBentryALTinterwordspacing
S.~Kassing and other, ``{Exploring the "Internet from Space" with Hypatia},'' \emph{IMC}, 2020. [Online]. Available: \url{https://doi.org/10.1145/3419394.3423635}
\BIBentrySTDinterwordspacing

\bibitem{etsi-kepler}
\BIBentryALTinterwordspacing
``{ETSI Kepler Web Reference Page},'' [Accessed 2022-Apr-30]. [Online]. Available: \url{https://www.etsi.org/deliver/etsi_tr/102500_102599/102505/01.02.01_60/tr_102505v010201p.pdf}
\BIBentrySTDinterwordspacing

\bibitem{web-moz-ttfb}
\BIBentryALTinterwordspacing
``{MDN Web Docs - Time to first byte},'' [Accessed 2022-Apr-30]. [Online]. Available: \url{https://developer.mozilla.org/en-US/docs/Glossary/time_to_first_byte}
\BIBentrySTDinterwordspacing

\bibitem{web-moz-fcp}
\BIBentryALTinterwordspacing
``{MDN Web Docs - First contentful paint},'' [Accessed 2022-Apr-30]. [Online]. Available: \url{https://developer.mozilla.org/en-US/docs/Glossary/First_contentful_paint}
\BIBentrySTDinterwordspacing

\bibitem{web-moz-plt}
\BIBentryALTinterwordspacing
``{MDN Web Docs - Page load time},'' [Accessed 2022-Apr-30]. [Online]. Available: \url{https://developer.mozilla.org/en-US/docs/Glossary/Page_load_time}
\BIBentrySTDinterwordspacing

\bibitem{ietf-masque-connect-udp}
\BIBentryALTinterwordspacing
D.~Schinazi, ``{UDP Proxying Support for HTTP},'' IETF, Internet-Draft draft-ietf-masque-connect-udp-09, Oct. 2021, work in Progress. [Online]. Available: \url{https://datatracker.ietf.org/doc/draft-ietf-masque-connect-udp/}
\BIBentrySTDinterwordspacing

\bibitem{kuhn2021}
\BIBentryALTinterwordspacing
N.~Kuhn \emph{et~al.}, ``{Evaluating BDP FRAME extension for QUIC},'' [Accessed 2022-Apr-30]. [Online]. Available: \url{https://arxiv.org/abs/2112.05450}
\BIBentrySTDinterwordspacing

\bibitem{quic-perf-sat-iwcmc}
\BIBentryALTinterwordspacing
S.~Yang \emph{et~al.}, ``{Performance Analysis of QUIC Protocol in Integrated Satellites and Terrestrial Networks},'' in \emph{IWCMC}, 2018. [Online]. Available: \url{http://doi.org/10.1109/IWCMC.2018.8450388}
\BIBentrySTDinterwordspacing

\bibitem{zhang2019}
\BIBentryALTinterwordspacing
H.~Zhang \emph{et~al.}, ``{How Quick Is QUIC in Satellite Networks},'' \emph{CSPS}, 2019. [Online]. Available: \url{https://doi.org/10.1007/978-981-10-6571-2_47}
\BIBentrySTDinterwordspacing

\bibitem{gquic_satcom}
\BIBentryALTinterwordspacing
L.~Thomas \emph{et~al.}, ``Google quic performance over a public satcom access,'' \emph{Int. J. Satell. Commun. Netw.}, vol.~37, pp. 601--611, 2019. [Online]. Available: \url{https://doi.org/10.1002/SAT.1301}
\BIBentrySTDinterwordspacing

\bibitem{satellite-quic}
\BIBentryALTinterwordspacing
J.~{Deutschmann} \emph{et~al.}, ``Satellite internet performance measurements,'' in \emph{2019 International Conference on Networked Systems (NetSys)}, 2019, pp. 1--4. [Online]. Available: \url{https://doi.org/10.1109/NetSys.2019.8854494}
\BIBentrySTDinterwordspacing

\bibitem{kuhn2020}
\BIBentryALTinterwordspacing
N.~Kuhn \emph{et~al.}, ``{QUIC: Opportunities and threats in SATCOM},'' \emph{ASMS/SPSC}, 2020. [Online]. Available: \url{https://doi.org/10.1109/ASMS/SPSC48805.2020.9268814}
\BIBentrySTDinterwordspacing

\bibitem{quic.acks.satcom}
\BIBentryALTinterwordspacing
A.~Custura \emph{et~al.}, ``{Impact of Acknowledgements using IETF QUIC on Satellite Performance},'' \emph{ASMS/SPSC}, 2020. [Online]. Available: \url{https://doi.org/10.1109/ASMS/SPSC48805.2020.9268894}
\BIBentrySTDinterwordspacing

\bibitem{rfc3449}
\BIBentryALTinterwordspacing
M.~Sooriyabandara \emph{et~al.}, ``{TCP Performance Implications of Network Path Asymmetry},'' RFC 3449. [Online]. Available: \url{https://www.rfc-editor.org/info/rfc3449}
\BIBentrySTDinterwordspacing

\bibitem{quic_sat_bbr}
\BIBentryALTinterwordspacing
Y.~{Wang} \emph{et~al.}, ``{Performance Evaluation of QUIC with BBR in Satellite Internet},'' in \emph{WiSEE}, 2018. [Online]. Available: \url{https://doi.org/10.1109/WiSEE.2018.8637347}
\BIBentrySTDinterwordspacing

\bibitem{long-look-at-quic}
\BIBentryALTinterwordspacing
A.~M. Kakhki \emph{et~al.}, ``{Taking a Long Look at QUIC: An Approach for Rigorous Evaluation of Rapidly Evolving Transport Protocols},'' \emph{IMC}, 2017. [Online]. Available: \url{https://doi.org/10.1145/3131365.3131368}
\BIBentrySTDinterwordspacing

\bibitem{http-over-quic}
\BIBentryALTinterwordspacing
G.~Carlucci \emph{et~al.}, ``{HTTP over UDP: An Experimental Investigation of QUIC},'' in \emph{SAC}, 2015. [Online]. Available: \url{https://doi.org/10.1145/2695664.2695706}
\BIBentrySTDinterwordspacing

\bibitem{when-quic-meets-tcp}
\BIBentryALTinterwordspacing
Y.~{Yu} \emph{et~al.}, ``{When QUIC meets TCP: An experimental study},'' in \emph{IPCCC}, 2017. [Online]. Available: \url{https://doi.org/10.1109/PCCC.2017.8280429}
\BIBentrySTDinterwordspacing

\bibitem{cook2017}
\BIBentryALTinterwordspacing
S.~Cook \emph{et~al.}, ``{QUIC: Better for what and for whom?}'' in \emph{ICC}, 2017. [Online]. Available: \url{https://doi.org/10.1109/ICC.2017.7997281}
\BIBentrySTDinterwordspacing

\bibitem{nottingham2018}
\BIBentryALTinterwordspacing
M.~Nottingham, ``{What's Happening with QUIC},'' [Accessed 2022-Apr-30]. [Online]. Available: \url{https://www.ietf.org/blog/whats-happening-quic/}
\BIBentrySTDinterwordspacing

\bibitem{saif2021}
\BIBentryALTinterwordspacing
D.~Saif \emph{et~al.}, ``{An Early Benchmark of Quality of Experience Between HTTP/2 and HTTP/3 using Lighthouse},'' \emph{ICC}, 2021. [Online]. Available: \url{https://doi.org/10.1109/ICC42927.2021.9500258}
\BIBentrySTDinterwordspacing

\bibitem{yu2021}
\BIBentryALTinterwordspacing
A.~Yu and T.~A. Benson, ``{Dissecting Performance of Production QUIC},'' \emph{WWW}, 2021. [Online]. Available: \url{https://doi.org/10.1145/3442381.3450103}
\BIBentrySTDinterwordspacing

\end{thebibliography}


\end{document}